\documentclass[prl,aps,twocolumn,superscriptaddress]{revtex4-1}
\usepackage{graphicx,color}
\usepackage{amsthm}
\usepackage{amsfonts}
\usepackage{algorithmic}
\usepackage{enumerate}
\usepackage{latexsym}
\usepackage{amsmath}
\usepackage{amssymb}
\usepackage[colorlinks=true,citecolor=blue,linkcolor=blue]{hyperref}
\def\avg#1{\left\langle#1\right\rangle}

\emergencystretch=\maxdimen
\begin{document}

\title{Antiferromagnetically ordered Mott insulator and $d+id$ superconductivity in twisted bilayer graphene: A quantum Monte carlo study}
\author{Tongyun Huang}
\affiliation{Department of Physics, Beijing Normal University, Beijing 100875, China\\}
\author{Lufeng Zhang}
\affiliation{Department of Physics, Beijing Normal University, Beijing 100875, China\\}

\author{Tianxing Ma}
\email{txma@bnu.edu.cn}
\affiliation{Department of Physics, Beijing Normal University, Beijing 100875, China\\}

\begin{abstract}
Using exact quantum Monte Carlo method, we examine the recent novel electronic states seen in magic-angle graphene superlattices. From the Hubbard model on a double-layer honeycomb lattice with a rotation angle $\theta=1.08^{\circ}$, we reveal that an antiferromagnetically ordered Mott insulator emerges beyond a critical $U_c$ at half filling, and with a small doping, the pairing with $d+id$ symmetry dominates over other pairings at low temperature. The effective $d+id$ pairing interaction strongly increase as the on-site Coulomb interaction increases, indicating that the superconductivity is driven by electron-electron correlation.   
Our non-biased numerical results demonstrate that the twisted bilayer graphene share the similar superconducting mechanism of high temperature superconductors, which is a new and ideal platform for further investigating the strongly correlated phenomena.


\end{abstract}

\maketitle
\noindent
\underline{\it Introduction}:
In past decades, studies on the exotic correlated electronic phases in graphene open up a new frontier in condensed matter physics\cite{Novoselov2005,Zhang2005,Peres2010,DasSarma2011}. Among these exciting research fields,
enormous theoretical proposals
have been made on engineering possible novel superconductivity (SC) in graphene \cite{PhysRevB.65.212505,PhysRevB.75.134512,PhysRevB.81.085431,PhysRevLett.98.146801,PhysRevLett.100.246808,PhysRevB.84.121410,Nandkishore2012,
PhysRevB.90.245114,PhysRevB.89.144501,PhysRevB.81.224505,PhysRevB.89.144501,PhysRevB.84.121410,RevModPhys.83.1057,RevModPhys.82.3045,PhysRevLett.108.147001,Profeta2012,Levitov2012}.
Previous studies suggest that it is a very challenging problem to induce SC near the charge neutrality point in graphene as the density of state (DOS) is rather low due to its Dirac-cone band, and if heavily doped, unconventional SC with different pairing symmetry is proposed, while the doping level is beyond current experimental capacity\cite{PhysRevB.84.121410,Nandkishore2012,
PhysRevB.90.245114}. Most recently, a series of breakthrough experiments on magic-angle graphene superlattices
  have triggered great excitement\cite{Cao2018A,Cao2018B}. By arranging two layers of atom-thick graphenes twisted at a narrow range of particular magic angle, the band structure of such twisted bilayer graphene (TBG) becomes nearly flat, and the Fermi velocity drops to zero in the vicinity of the Fermi energy.
Intriguingly, this system is interpreted as a correlated Mott insulator at half filling\cite{Cao2018A}, and when a few extra charge carriers are doped in, the insulator turns into a superconductor at
1.7 K with charge carriers density $\sim10^{11}/cm^{-2}$\cite{Cao2018B}. 

Regarding this ultra low doping density, the transition temperature of 1.7 K is remarkably high, and the SC is suggested to be
originated from electron correlation, which has a striking similar trend as that in doped cuprates\cite{Bednorz1986}, heavy-fermion\cite{Steglich1979}, iron-based\cite{Kamihara2008} and organic superconductors\cite{Jerome1980}.
Thus, the realization of unconventional SC in TBG provides a relatively simple and more importantly, highly tunable and realistic platform for studying correlated electron physics,
especially, which holds promise for several long standing problems, for example, the understanding of unconventional SC, and also may prove to be a significant step in the searching for room-temperature superconductors\cite{Cao2018A,Cao2018B}.
Moreover, the vicinity between various magnetic orders and SC in high temperature superconductors is one of the most notorious issues.
These problems, are the biggest challenge of condensed matter physics\cite{RevModPhys.84.1383}, and the TBG, may provide an intriguing route to study the largely unknown physics.

\begin{figure}[tbp]
\includegraphics[scale=0.25]{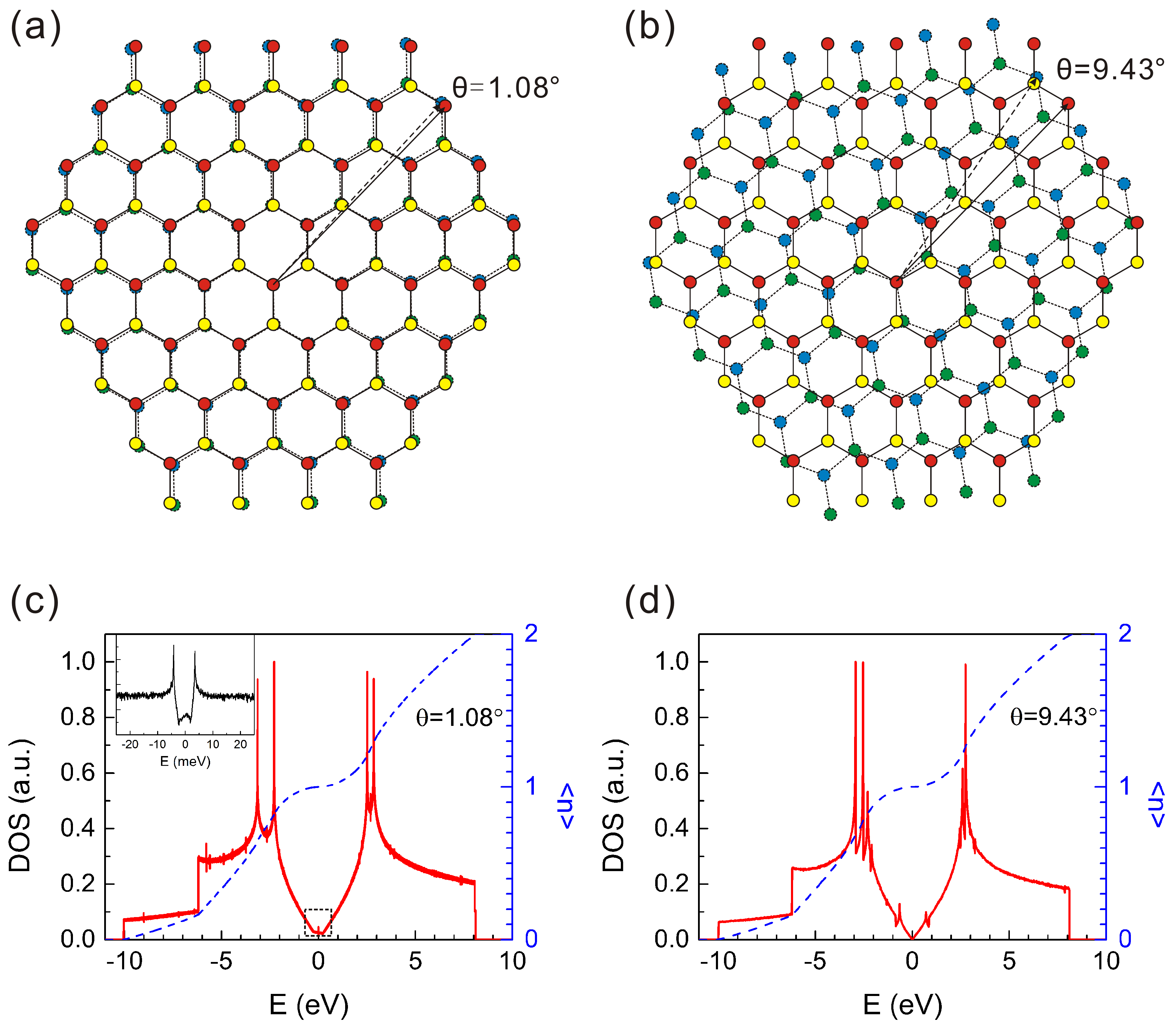}
\caption{(Color online) (a) Sketch of twist bilayer graphene with double-48 sites at $\theta=1.08^{\circ}$ and (b) $\theta=9.43^{\circ}$; (c) The DOS as function of energy with $\theta=1.08^{\circ}$ and (d) $\theta=9.43^{\circ}$.}
\label{Fig:Sketch}
\end{figure}

However, the nature of the superconducting phase and the correlated insulating state in TBG are under very active debate\cite{Cao2018A,Cao2018B}. Especially, to establish the mechanism and the paring symmetry for the observed SC are among the central theoretical challenges, while different pairing symmetries by various theoretical methods have been proposed\cite{Cao2018B,arxiv1803.08057,arxiv1803.11190,arxiv1804.00627,arxiv1804.03162,arxiv1803.04400}.
To win these great challenges, using unbiased numerical techniques is believed to be the only opportunity as Hartree-Fork-type approaches are biased if the electronic correlation dominates in the system. In current work, we are aiming to identify the Mott physics and the pairing symmetry in TBG by using exact quantum Monte Carlo (QMC) method. Here, our non-biased numerical results almost recover all the novel electronic states seen in TBG where the existence of SC close to an antiferromagnetically ordered Mott insulator\cite{Cao2018A,Cao2018B}, which is a hallmark of doped cuprates and other unconventional superconductors\cite{RevModPhys.84.1383}.
Our study marks the first step in dealing with similar fundamental issues for vertically twist stacked correlated materials, which may open a new direction for the investigation of strongly correlated phases of matter.

\noindent
\underline{\it Model and method}:
The sketches for TBG with rotation angle $\theta=1.08$ and $\theta=9.43$ between the layers have been shown in Fig.\ref{Fig:Sketch} (a) and (b) respectively, and $\theta$, which are related to $(m,n)$ by $\cos \theta =\frac{{{m}^{2}}+{{n}^{2}}+4mn}{2\left( {{m}^{2}}+{{n}^{2}}+mn \right)}$, coincide with the value of (31,30) and (4,3) for the fully optimized geometries of TBG in Ref.\cite{PhysRevB.90.155451}. The parameters $(m,n)$
are corresponding to the basis vector $\mathbf{v}_{1} = {m}\mathbf{a}_{1} +{n}\mathbf{a}_{2}$ of the first layer and $\mathbf{v}_{2} = {n}\mathbf{a}_{1} + {m}\mathbf{a}_{2}$ of the second layer for the non-rotating bilayer graphene, and they merge after one layer rotates the angle $\theta$.
Here, $\mathbf{a}_{1}$ and $\mathbf{a}_{2}$ are the lattice vectors of each sublattice.
In that geometry, each lattice consists of two layers, and each layer includes two interpenetrating triangular sublattices {\it with hexagonal shape such that it preserves most geometric symmetries of graphene}\cite{PhysRevB.84.121410,PhysRevB.90.245114,doi:10.1063/1.3485059}. In each sublattice, the total number of unit cells is $3L^2$ and the total number of lattice sites is $N_s$=2$\times$2$\times$3$L^2$. According to Ref.\cite{PhysRevB.90.155451}, there is a critical angle $\theta_c=5^{\circ}$, and bellow which the Fermi velocity decreases dramatically toward zero to cause flat bands at the Fermi level. In Fig.\ref{Fig:Sketch} (c), it is clear to see there is a Van Hove singularity (VHS) at half filling with $\theta=1.08^{\circ}$, agreeing with the experimental reports\cite{Cao2018B}. For comparation, the DOS at $\theta=9.43^{\circ}$ are also shown in Fig.\ref{Fig:Sketch} (d), which has a splitting of VHS in higher energy.

Including the electronic correlation, the Hamiltonian for the twisted bilayer honeycomb lattice reads\cite{PhysRevLett.99.256802,PhysRevB.96.155416,PhysRevB.82.121407}
\begin{eqnarray}
H=&-&t\sum_{l\left\langle i,j\right\rangle \sigma}(a^{\dagger}_{li\sigma}b_{lj\sigma}+b^{\dagger}_{li\sigma}a_{lj\sigma})\nonumber\\
&-&\sum_{i,j,l\neq l'\sigma}t_{ij}(a^{\dagger}_{li\sigma}a_{l'j\sigma}+a^{\dagger}_{li\sigma}b_{l'j\sigma}+b^{\dagger}_{li\sigma}a_{l'j\sigma}+b^{\dagger}_{li\sigma}b_{l'i\sigma}) \nonumber\\
&+&\mu\sum_{i,l,\sigma}(a^{\dagger}_{li\sigma}a_{li\sigma}+b^{\dagger}_{li\sigma}b_{li\sigma}) \nonumber\\
&+&U\sum_{i,l}(n_{lai\uparrow}n_{lai\downarrow}+n_{lbi\uparrow}n_{lbi\downarrow}), \label{model}
\end{eqnarray}
where $a_{li\sigma}$ ($a_{li\sigma}^{\dag}$) annihilates (creates) electrons
at site $\mathbf{R}^{a}_{li}$ of $l$ layer with spin $\sigma$ ($\sigma$=$\uparrow,\downarrow$)
on sublattice A, $b_{li\sigma}$ ($b_{li\sigma}^{\dag}$)
acts similar but on sublattice B,
$n_{lai\sigma}=a_{li\sigma}^{\dagger}a_{li\sigma}$ and
$n_{lbi\sigma}=b_{li\sigma}^{\dagger}b_{li\sigma}$.
$t\approx 2.7eV$ is the nearest-neighbor (NN) hopping integral, $\mu$ is the chemical potential and $U$ denotes the on-site
Hubbard interaction. In the following, we will take $t$ as the unit.
The interlayer hopping energy between sites $\mathbf{R}_{1i}$ and $\mathbf{R}_{2j}$ is
\begin{eqnarray}
t_{ij}=t_c e^{-[(|\mathbf{R}^{d}_{1i}-\mathbf{R}^{d'}_{2j}|)-d_0]/\xi}
\label{tc}
\end{eqnarray}
where the parameters are set as $t_c=-0.17$, $d_0=0.335$ nm, and $\xi=0.0453$ nm\cite{PhysRevB.96.155416}. Here $d_0$ indicates sites on sublattice A or B.
The interlayer hopping is considered over all sites in the geometry, and it decreases exponentially with the distance $|\mathbf{R}^{d}_{1i}-\mathbf{R}^{d'}_{2j}|$. Here, $\mathbf{R}^{d'}_{2j}=(\mathbf{R}^{d}_{2jx}cos\theta,  \mathbf{R}^{d}_{2jy}sin\theta)$ is the rotated position of $\mathbf{R}^{d}_{2j}$.

Our simulations are mostly performed on lattices of $N_s$=192 sites ($L$=4) with periodic boundary
conditions. To make the finite-size scaling analysis, lattices with $L$=2,3,4,5,6 are also simulated.
The basic strategy of the finite temperature determinant Monte Carlo (DQMC) method is to express
the partition function as a high-dimensional integral over a set of random auxiliary fields. The integral is then accomplished by
Monte Carlo techniques. In our simulations, 8 000 sweeps were used to equilibrate the system, and an
additional 10 000$\sim$200 000 sweeps were then made, each of which generated a
measurement. These measurements were split into ten bins which provide the
basis of coarse-grain averages and errors were estimated based on standard
deviations from the average. In order to assess our results and
their accuracy with respect to the infamous sign problem as the particle-hole
symmetry is broken, a very careful analysis on the average of sign is illustrated, and the results by constrained-path Monte Carlo (CPQMC) method are also present, where the sign problem is eliminated by the constrained-path approximation\cite{PhysRevB.84.121410,PhysRevLett.74.3652,PhysRevLett.78.4486}.
%

\begin{figure}[tbp]
\includegraphics[scale=0.42]{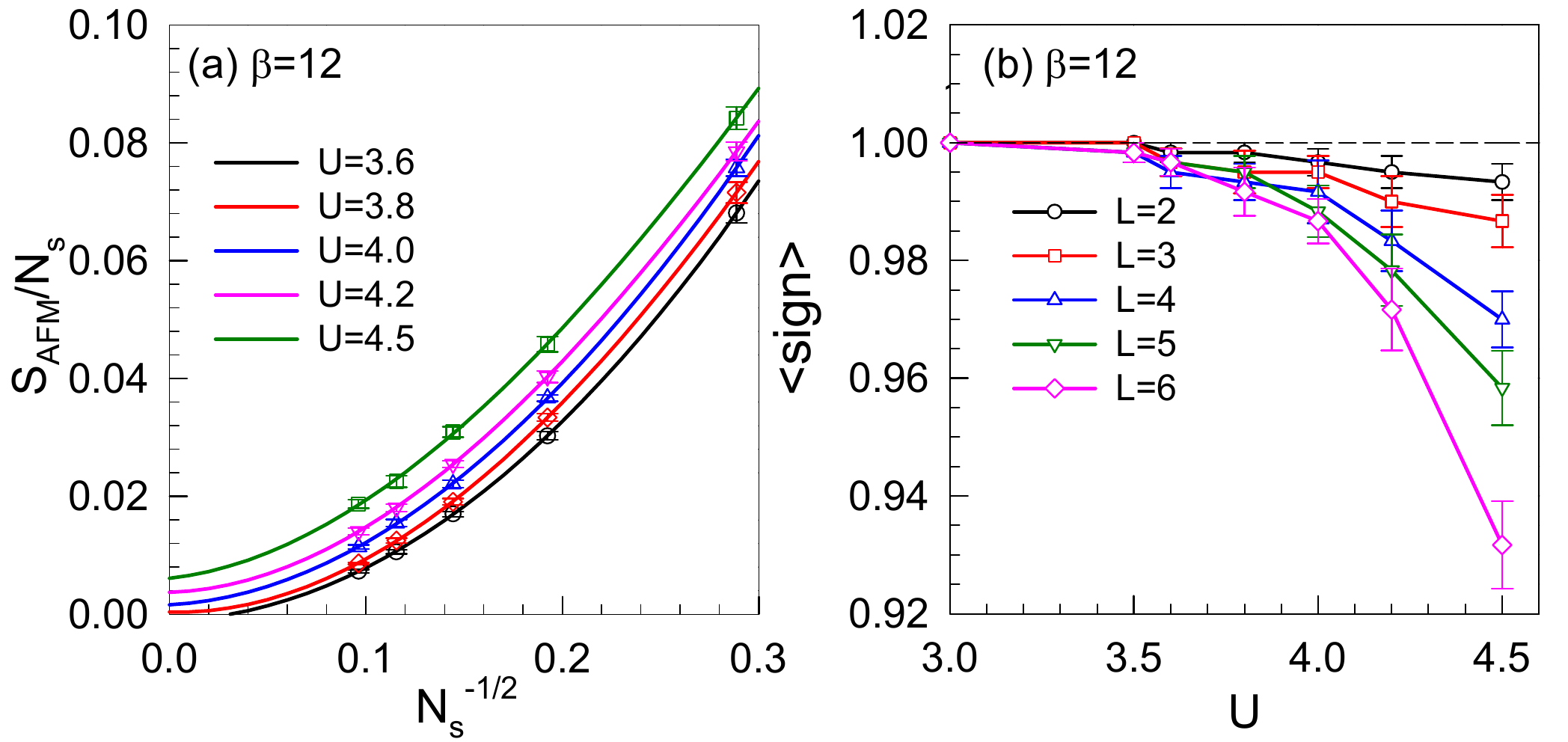}
\caption{(Color online) (a) Scaling behavior of the normalized AFM spin structure factor $S_{AFM}/N_{s}$ for different values of $U$ at $\beta=12$. Solid lines are fit of the third-order polynomial in $1/\sqrt{N_s}$. (b) The corresponding $\avg{sign}$ at $\beta=12$.
} \label{Fig:Spin}
\end{figure}

\noindent
\underline{\it Results and discussion}---
As magnetic order plays a key role in the superconducting mechanism
of electronic correlated systems, we first study the antiferromagnetic (AFM) spin structure factor
\begin{equation}
S_{AFM}=\frac{1}{N_s}\langle [\sum_{{lr}}({\hat{S}^{z}_{lar}}-{\hat{S}^{z}_{lbr}})]^{2}  \rangle,
\label{spin}
\end{equation}
which indicates the onset of long-range AFM order if $\lim_{N_s\rightarrow\infty}(S_{AFM}/N_s)>$0. Here, ${\hat{S}^{z}_{lar}}$(${\hat{S}^{z}_{lbr}}$) is the $z$ component spin operator on A (B) sublattice of layer $l$. $S_{AFM}$ for different interactions
are calculated on lattices with $L=2,3,4,5,6$, and are extrapolated to the thermodynamic limit using polynomial functions in $1/\sqrt{N_s}$. As that shown in Fig. \ref{Fig:Spin}(a), one can deduce that the critical $U_c$, where the AFM long range order develops, is around $3.8$.
The average sign, $\avg{sign}$ with 10 1000 runs are shown in Fig. \ref{Fig:Spin}(b), which is larger than 0.92 at $U$ up to $4.5$ and $N_s$ up to $432$ for the lowest temperature we reached. In order to obtain the same quality of data as $\avg{sign} \simeq 1$, much longer runs are necessary to compensate the fluctuations. Indeed, we can estimate that the runs need to be stretched\cite{PhysRevD.24.2278,SANTOS2003,PhysRevB.94.075106} by a factor on the order of  $\avg{sign}^{-2}$. In our simulations, especially in further Figs. 3, 5 and 6 where the sign problem is much worse, we have increased measurement from 10 000 to 200 000 times to compensate the fluctuations, and thus the results for current parameters are reliable. 

\begin{figure}[tbp]
\includegraphics[scale=0.4]{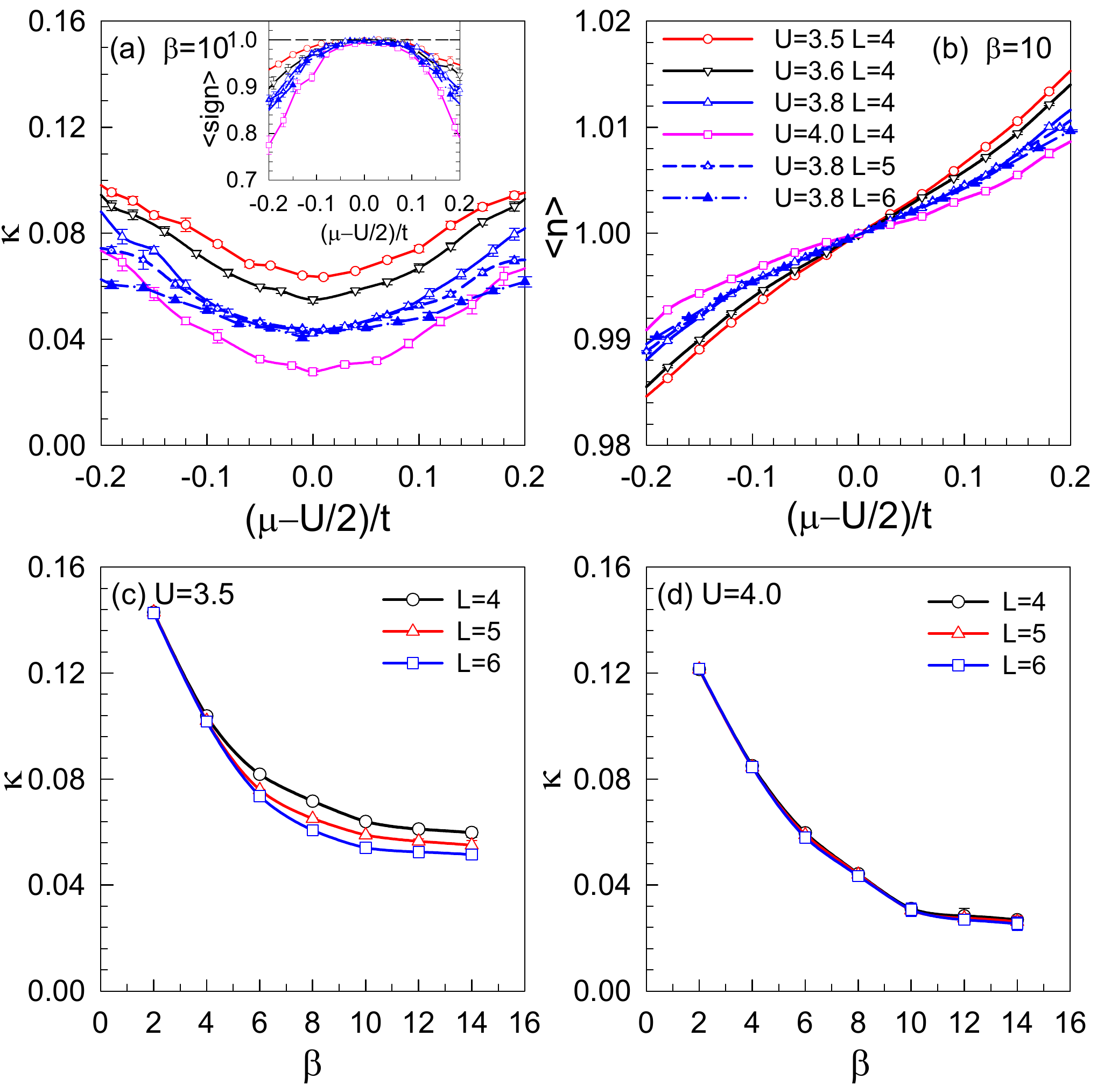}
\caption{(Color online)
(a) Charge compressibility $\kappa$ and (b) electron filling $\avg{n}$  versus $\mu$ at $\beta=10$ for several interaction strengthes. The dependency of lattice linear size $L$ for $\kappa$ at $U =3.5$ (c) and $U=4.0$ (d). Inset: the corresponding $\avg{sign}$ for different values of $U$ and $L$ at $\beta=10$.
} \label{Fig:Kappa}
\end{figure}

One electronic state of high interest is the Mott-like insulator\cite{Cao2018A}.
For single layer honeycomb lattice, it exhibits a charge (Mott) excitation gap at sufficiently large $U$ at half-filling\cite{PhysRevLett.120.116601,PhysRevX.6.011029}, and the
single-particle gap can be used to establish the existence of the Mott insulator.
Basically the single-particle gap should be extracted from DOS, and here we extract the information of the gap by examining the behavior of charge compressibility $\kappa(\mu)=d\avg{n(\mu)}/d\mu$ at the Fermi level. 
The threshold of $\kappa$ is finite on finite lattices at non-zero temperature, and vanishing compressibility will overestimate the critical coupling strength due to temperature broadening effects\cite{Lee2007}. After analyzing the effect of finite $T$ in the noninteracting limit, we take $\kappa\sim 0.04$ as an appropriate threshold to distinguish between gapped ($\kappa< 0.04$) and gapless($\kappa>0.04$) system\cite{PhysRevLett.120.116601}.
Results for $\kappa(\mu)$ evaluated at inverse temperature $\beta=10$ are depicted in Fig.\ref{Fig:Kappa} (a) with various $U$. And we can also tell this from Fig.\ref{Fig:Kappa} (b) while $\avg{n(\mu)}$ converges faster than $\kappa$ vanishes.
Fig.\ref{Fig:Kappa} (a) suggests that the system becomes incompressible at $U_c\sim 3.8$, and combine results shown in Fig.\ref{Fig:Spin}, we identify that the state at half filling with $U>U_c$ is an antiferromagneticlly ordered Mott insulating state.
In the insert of Fig.\ref{Fig:Kappa} (a), we can see that $\avg{sign}$ is mostly larger than 0.75 for $\kappa$ at $\beta=10$ with $L=4,5,6$ and $U\leq 4.0$.
We further compare the impact of lattice sizes on $\kappa$ and $\avg{n}$ for $L=4,5,6$ with $U = 3.5, 3.8$ and $4.0$, in the metallic and insulating region, respectively. Our calculations show that, in the metallic region, Fig.\ref{Fig:Kappa} (c), it shows a stronger size dependence, while in the insulating region, Fig.\ref{Fig:Kappa} (d), it is nearly free of finite size effect.

To investigate the superconducting property of TBG, we compute the pairing
susceptibility
\begin{equation}
P_{\alpha}=\frac{1}{N_s}\sum_{l,i,j}\int_{0}^{\beta }d\tau \langle \Delta
_{l,\alpha }^{\dagger }(i,\tau)\Delta _{l,\alpha }^{\phantom{\dagger}%
}(j,0)\rangle,
\label{Pa}
\end{equation}
where $\alpha$ stands for the pairing symmetry.
Due to the constraint of on-site Hubbard interaction in
Eq.\ref{model}, pairing between two sublattices is favored and the corresponding
order parameter $\Delta _{l\alpha }^{\dagger }(i)$\ is
\begin{eqnarray}
\Delta _{l\alpha }^{\dagger }(i)\ =\sum_{\bf l}f_{\alpha}^{\dagger}
(\delta_{\bf l})(a_{{li}\uparrow }b_{{li+\delta_{\bf l}}\downarrow }-
a_{{li}\downarrow}b_{{li+\delta_{\bf l}}\uparrow })^{\dagger},
\label{Ca}
\end{eqnarray}
with $f_{\alpha}(\bf{\delta}_{\bf l})$ being the form factor of pairing function.
In order to extract the intrinsic pairing interaction in finite system, one should subtract from $P
_{\alpha}$ its uncorrelated single-particle contribution $\widetilde{P}
_{\alpha}$, which is achieved by replacing $\langle
a_{{li}\downarrow }^{\dag }a_{{lj}\downarrow }b_{i+\delta_{\bf l}\uparrow}^{\dag}
b_{j+\delta_{\bf l'}\uparrow}\rangle $ in Eq. (\ref{Ca}) with $\langle a_{{i}\downarrow }^{\dag
}a_{{j}\downarrow }\rangle \langle b_{i+\delta_{\bf l}\uparrow }^{\dag }
b_{j+\delta_{\bf l'}\uparrow }\rangle $, and we have the intrinsic pairing interaction  ${\bf P_{\alpha}}=P_{\alpha}-\widetilde{P}_{\alpha}$.

\begin{figure}[tbp]
\includegraphics[scale=0.45]{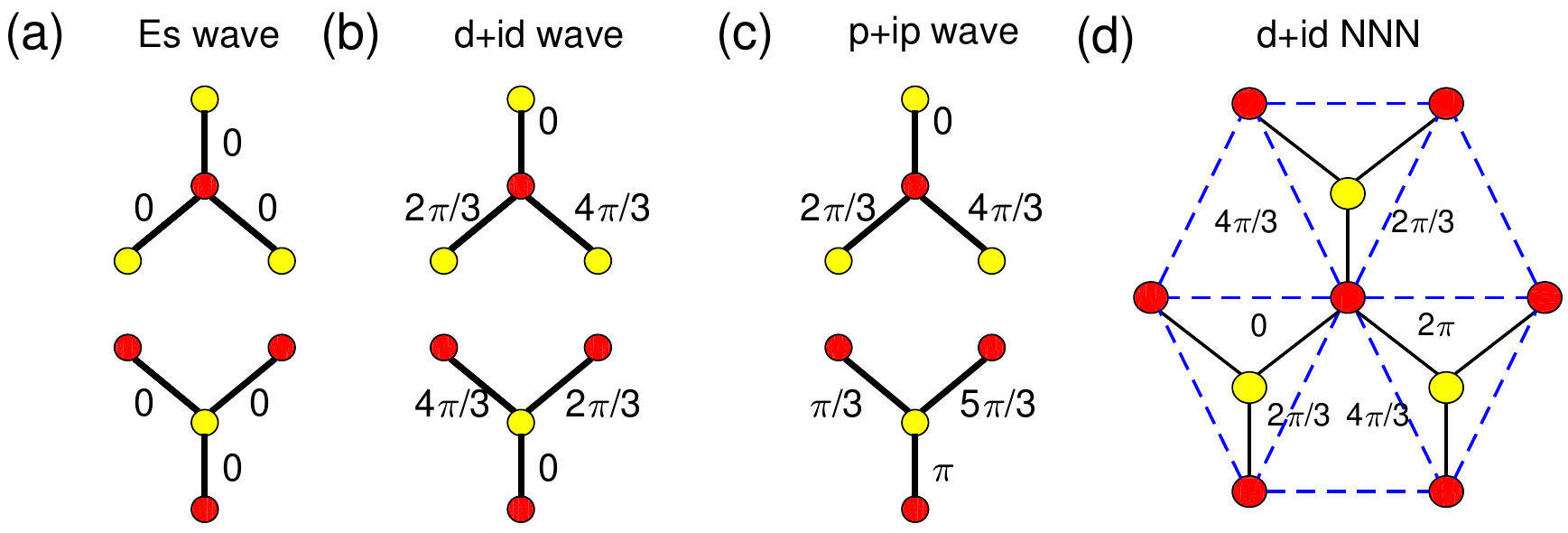}
\caption{(Color online) Phases of the pairing symmetries of (a) extensive $S$ (b) $d+id$ (c) $p+ip$ and (d) $d+id$ wave with next nearest neighbour.}
\label{Fig:Pairing}
\end{figure}

In Eq.\ref{Ca}, the vectors $\bf{\delta_{\bf l}}$ $({\bf l}=1,2,3)$ denote the NN inter sublattice connections sketched in Fig.~\ref{Fig:Pairing}.
Considering the special structure of honeycomb lattice, the possible pairing symmetries are given by (a) extensive $S$ ($ES$) (b) $d+id$ and (c) $p+ip$ wave\cite{PhysRevB.84.121410,PhysRevB.90.245114,0295-5075-111-4-47003}, whose form factors are illustrated in Fig.~\ref{Fig:Pairing}. These different pairing symmetries are distinguished by different phase shifts upon $\pi/3$ or $2\pi/3$ rotations. The singlet $ES$ wave and NN-bond $d+id$ pairing has the form factor
\begin{eqnarray}
\ f_{ES}({\bf{\delta}_{l}})=1, ~{\bf{l}}=1,2,3\\
\ f_{d+id}({\bf{\delta}_{l}})=e^{i({\bf{l}}-1)
\frac{2\pi }{3}},~{\bf{l}}=1,2,3,
\end{eqnarray}
{for whatever sublattice A and B, while the NN-bond $f_{p+ip}$ is different for A and B sublattice, where 
\begin{eqnarray}
\ f_{p+ip}({\bf{\delta}_{al}})=e^{i({\bf l}-1)\frac{2\pi }{3}},~{\bf l}=1,2,3,
\end{eqnarray}
for A, and accordingly the phase on the same link for B, there is a $\pi$ phase shift,
$\ f_{p+ip}({\bf{\delta}_{bl}})=e^{i({\bf l}-1)\frac{2\pi }{3}+\pi}$.
We also studied longer range pairings by adding next nearest neighbour (NNN) bond pairing for $d+id$ wave symmetry,
which have the following form factors
\begin{eqnarray}
\ f_{d+id}({\bf{\delta}_{l}})=e^{i({\bf l}-1)\frac{2\pi }{3}}~{\bf l}=1,2,3...6.
\end{eqnarray}

\begin{figure}[tbp]
\includegraphics[scale=0.42]{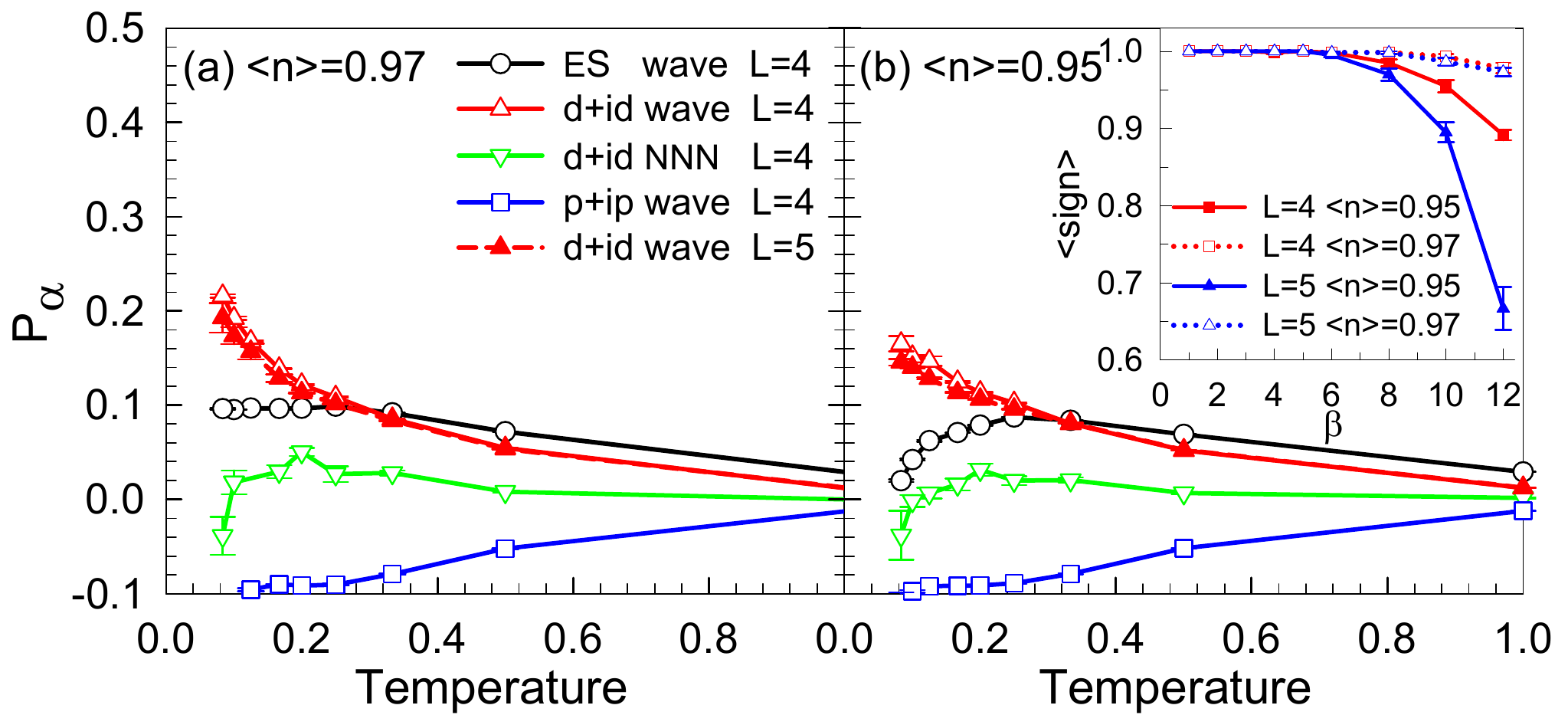}
\caption{(Color online) ${\bf P}_{\alpha}$ as a function of temperature at (a) $\avg{n}=0.97$ and (b) $\avg{n}=0.95$ for $U=3.0$. Inset: the temperature-dependent $\avg{sign}$ at $\avg{n}=0.95, 0.97$ with the corresponding $L$ for $U=3.0$.
}
\label{Fig:Pairingn}
\end{figure}

As it is expected that fermion systems with strong on-site
repulsion may exhibit SC induced by AFM spin fluctuations,
and from the behavior of magnetic correlation shown in Fig.~\ref{Fig:Spin},
it seems that the pair formation is possible through a similar mechanism
in TBG. In the following, we discuss the behavior of the effective pairing interaction in a very low doped region,
which is what the experiment has been performed\cite{Cao2018B}.

\begin{figure}[tbp]
\includegraphics[scale=0.42]{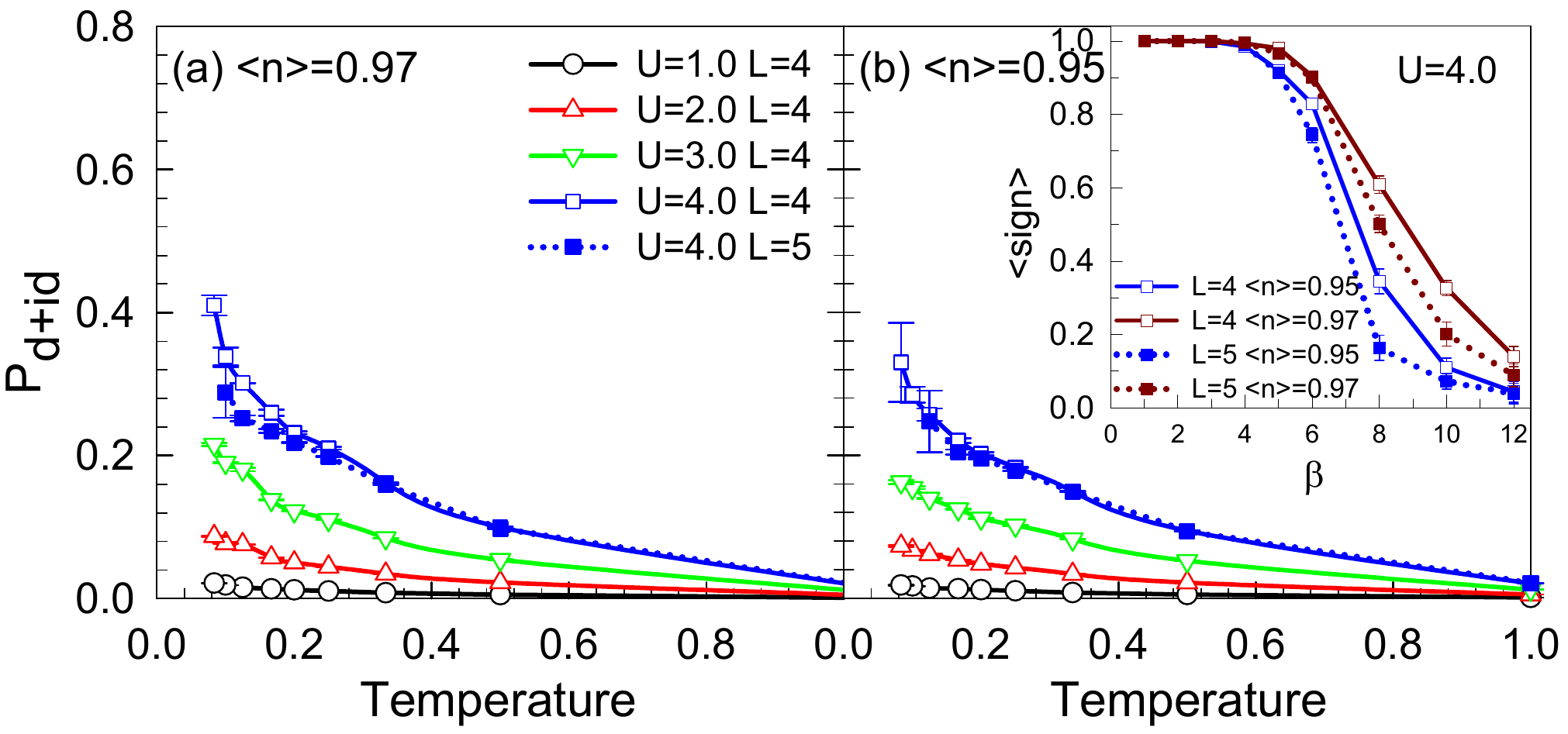}
\caption{(Color online) ${\bf P}_{d+id}$ as a function of temperature at (a) $\avg{n}=0.97$ and (b) $\avg{n}=0.95$ with different interaction strength. Inset: the temperature-dependent $\avg{sign}$ at $\avg{n}=0.95, 0.97$ with the corresponding $L$ for $U=4.0$.}
\label{Fig:PairingU}
\end{figure}

Fig.~\ref{Fig:Pairingn} shows the temperature dependence of ${\bf P}_{\alpha}$ for different pairing symmetries
at (a) $\avg{n}=0.97$ and (b) $\avg{n}=0.95$ with $U=3.0$. It is clearly to see that, the effective pairing interaction
with $d+id$ symmetry is always positive and increases with the lowering of temperature,
and it is almost independent on lattice size in high temperature region, and the dependence on lattice size is very weak in low temperature region.
Such a temperature dependence of ${\bf P}_{d+id}$ suggests effective attractions generated between electrons and the instability toward SC in the system at low temperatures for both $\avg{n}=0.97$ and $\avg{n}=0.95$. As for the other two pairing symmetries, $p+ip$ wave and $d+id$ NNN shown, our DQMC results yield negative effective pairing interactions, reflecting the fact that the realization of the $d+id$ symmetry at low temperatures will suppress other competing pairing channels.

Moreover, Fig. \ref{Fig:PairingU}(a) and(b) show that the effective pairing interaction for $d+id$ symmetry enhances with larger $U$. Especially, ${\bf P}_{d+id}$ tends to diverge in low temperature region as $U>3.0$, and the increasing $U$ tends to promote such diverge.  This demonstrates that the $d+id$ pairing SC is driven by strong electronic correlation.
In addition, $\avg{sign}$ is larger than 0.65 for $L=4,5,6$ at $U\leq3.0$ as shown in the insert of Fig.~\ref{Fig:Pairingn} (b). At a larger $U=4.0$ shown in the insert of Fig. \ref{Fig:PairingU}(b), $\avg{sign}$ is mostly larger than 0.2 as $\beta\leq8.0$ for $L=5$. For $U=4.0$ and $\beta>8.0$, the sign problem is worse while which is not important as the dominant pairing symmetry is robust on the temperature.

\begin{figure}[tbp]
\includegraphics[scale=0.45]{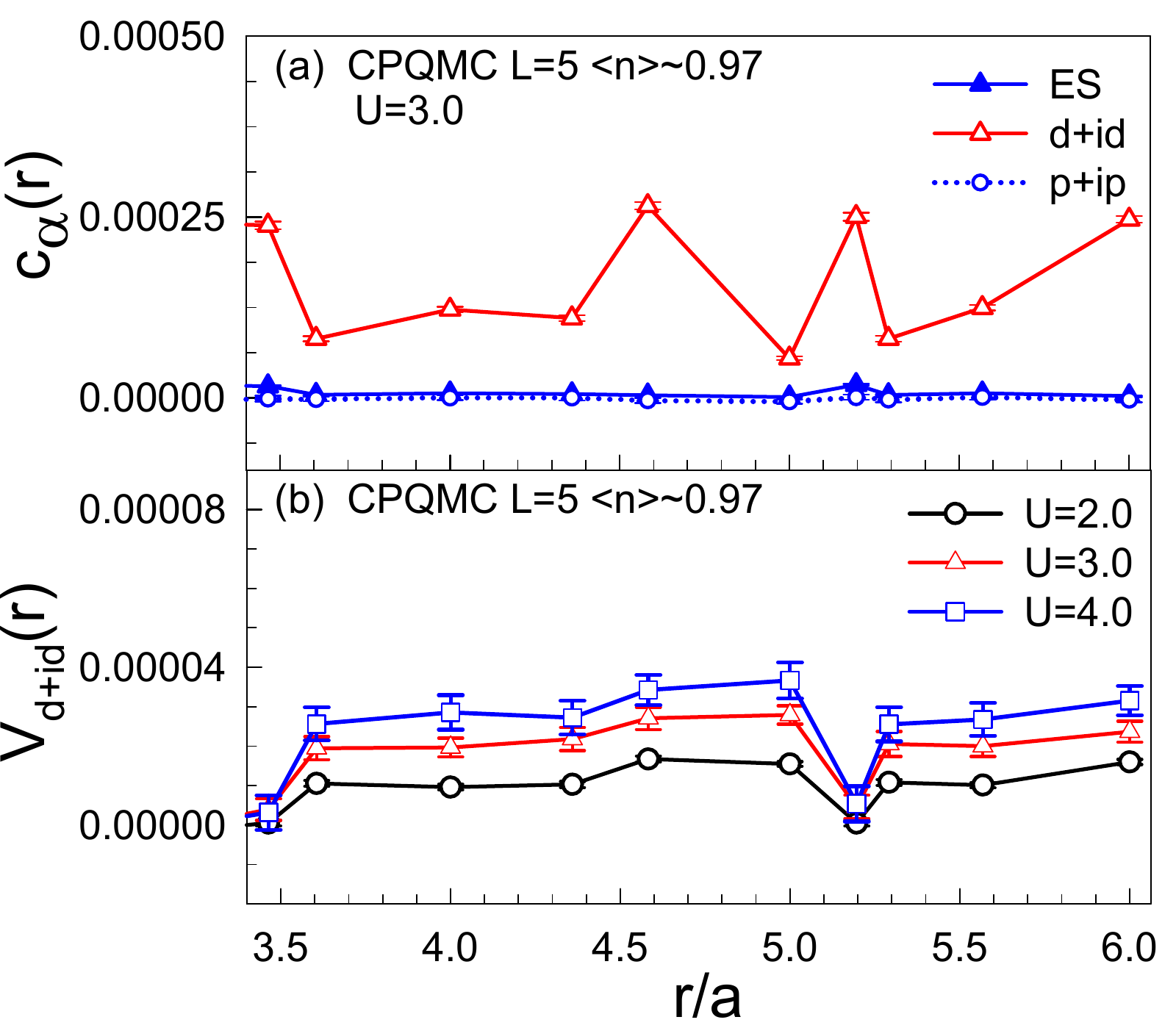}
\caption{(Color online)Pairing correction $C_{\alpha}$ as a function of distance $r$ for (a) different pairing symmetries with $U=3.0$, (b) the vertex contributions of $d+id$-wave with different $U$ at $\avg{n}\simeq0.97$ and $L=5$.}
\label{Fig:PairingCorrection}
\end{figure}

In general, to determine which pairing symmetry is dominant by numerical
calculation for finite size models, we had better to look at the long-range
part of the ground state pair-correlation function\cite{PhysRevLett.70.682,PhysRevLett.78.4486,PhysRevB.84.121410}, which could be achieved by the CPQMC method. In Fig. \ref{Fig:PairingCorrection} (a), the distance dependent pairing-pairing correlation at zero temperature,
$
C_{\alpha}(r)= \Sigma_l\langle \Delta
_{l\alpha }^{\dagger }(i)\Delta _{l\alpha }^{\phantom{\dagger}%
}(j)\rangle,
$
 is shown for $\avg{n}=145/150\simeq 0.97$. It is clear to see that the $C_{d+id}(r)$ is larger than $C_{ES}(r)$  and $C_{p+ip}(r)$  for all long-range distances between electron pairs. This reenforces our finding that the $d+id$ pairing symmetry dominates other pairing symmetries.  We also examined the vertex contributions  ${\bf V_{\alpha}}=C_{\alpha}-\widetilde{C}_{\alpha}$ in Fig. \ref{Fig:PairingCorrection} (b), which increase as the interactions increase, indicating the importance of electronic correlation in enhancing SC.

\noindent
\underline{\it Summary}---
 In summary, we study the spin correlation, the charge compressibility and the superconducting pairing symmetry in TBG by using exact QMC method. From a double-layer honeycomb lattice with a rotation angle $\theta$=1.08$^{\circ}$, we almost recover all the recent
 experimentally observed novel electronic states in TBG. At half filling, an antiferromagnetically ordered Mott insulator emerges beyond a critical $U_c$$\sim$3.8. With a finite doping, the pairing with $d+id$ symmetry dominates over other pairing symmetries, and it increases fast as the interaction increases, indicating that the SC is driven by strong electronic correlations. Our exact numerical results demonstrate that the TBG holds a very similar interaction driven phase diagram of doped cuprates and other high temperature superconductors, which may provide a new and ideal platform to the unified understanding of the superconducting mechanism in electronic correlated system.

\noindent
\underline{\it Acknowledgement} ---
This work were supported by NSFCs (Grant.~Nos. 11374034 and 11334012). We acknowledge the support of
HSCC of Beijing Normal University, and phase 2 of the Special Program
for Applied Research on Super Computation of the NSFC-Guangdong Joint
Fund.

\bibliography{reference}

\end{document}